\documentclass[aps,eqsecnum,showpacs]{revtex4}
\def\c#1{\setbox0=\hbox{#1}\ifdim\ht0=1ex\accent24 #1%
  \else{\ooalign{\hidewidth\char24\hidewidth\crcr\unhbox0}}\fi}
  
\usepackage{graphicx}    
\usepackage{color}       
\usepackage{subfigure}   

\begin{document}
\pagestyle{empty}
\tighten
\title{
          TRANSPORT IN TWO--DIMENSIONAL QUANTUM WIRES
          WITH POINT SCATTERERS. THE WIGNER--SMITH TIME DELAY
         }
\author{\bf Robert G\c{e}barowski}
\address{
            Instytut Fizyki, Politechnika Krakowska im. Tadeusza Ko\'sciuszki,
            ulica Podchor\c{a}\.zych 1,
            30-084 Krak\'ow, Poland.
           }
\date{\today}

\begin{abstract}
Electronic transport properties of the disordered
quantum wires are considered.
The disorder is introduced via impurities (point scatterers),
distributed uniformly over the two-dimensional strip, which represents
a model quantum wire. Incident electrons with a given energy are scattered on 
impurities and boundaries of the wire. 
The electron-electron interaction is neglected in the model.
In particular, the intermediate regime and the 
localization regime of transport are studied in more detail in terms
of the conductance and statistical properties of $S$-matrix ensemble 
for a given incident electron energy. 
The Wigner-Smith time delay
distribution obtained for the localization regime is compared with
the prediction for the scattering by a one-dimensional random potential. 
\end{abstract}

\pacs{72.20.Dp,05.45.+b,72.10.Bg.}

\maketitle

\section{INTRODUCTION}

Quantum transport through disordered mesoscopic systems and
its description by the Random Matrix Theory (RMT) have
been subject of an intense research for some time \cite{B97}. The
phenomenon of quantum scattering is universal --- it is
also encountered in other physical systems, like for example
in scattering of electromagnetic radiation by reflecting
obstacles.

In order to study properties of the quantum transport,
a two-dimensional  model
 of the disordered wire is considered.
The model, in which a number of point scatterers is randomly
distributed over a strip, is continuous and yet solvable ---
expressions for scattering matrix (the $S$-matrix) elements can be
given explicitly \cite{EGST96}.

Therefore, by means of varying the number of point scatterers
together with the length of the sample, it is possible to modify
properties of the electron transport through the wire in a well
controlled way. Hence one can observe various types of
the system behaviour ranging from the ballistic type of transport through
``diffusive-like'' scattering to the localization for a given
energy of incoming electrons \cite{JP95,GSZZ98}. The model allows to study
both macroscopic transport properties, like conductance, and
microscopic signatures of the scattering processes such as
eigenphase, $S$-matrix elements' statistics and time delay distributions.

The aim of this contribution is to 
identify and investigate various 
regimes of the transport in the model quantum wire. 
The results will be compared  with predictions of the standard Random
Matrix Theory of transport. The signatures of various regimes
of transport in the presence of the Time Reversal Symmetry will
be considered in order to reveal regimes of the universal behaviour of the model
and also deviations from that behaviour.

\section{THE MODEL}

A disordered quantum wire is modelled with a two-dimensional rectangular strip
of length $L$ and width $W$ ($W=\pi$ is taken) with hard walls \cite{EGST96,GSZZ98}. 
There is a finite number, $N$, of point scatterers randomly and uniformly 
distributed over the scattering region with coordinates $(x_j, y_j)$:

\begin{equation}
(x_j, y_j) \in (-L/2, L/2) \times (0, \pi), \qquad {\rm for} \qquad j=1, \ldots ,N
\label{xy}
\end{equation}

An electron can enter the strip either from the left
or from the right side. The number of channels, $M$, in the model wire is equal
to the integer part of length of the wavevector $\bf {k}$ of the incoming electron.
 The scattering in the strip on impurities and boundaries 
 is assumed to be elastic and the electron--electron
 interaction interaction is neglected. The hard wall 
 boundary condition means that the wavefunction vanishes on the boundaries
 for all $x$:
\begin{equation}
 \psi(x,0) = \psi(x,W) = 0.
\label{dirichlet}
\end{equation}

\begin{figure}[h]
  \includegraphics[scale=0.45]{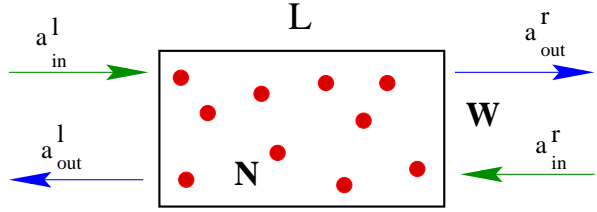}
  \caption{This illustrates a two-dimensional strip model of a quantum wire
  with $N$ point scatterers. The length of the strip is $L$ and the width is $W$
  It has been taken $W=\pi$ in order to have the number of open channels $M = [k]$
  (integer part of the incident electrons' wavenumber $k$; 
   the electrons' energy $E = k^2$ in applied units here).
          }
  \label{model}	  	  
\end{figure}

Figure~(\ref{model}) illustrates considered theoretical model of the quantum wire.

The scattering matrix $S$ relates incoming waves, denoted by
$M$-component vectors $\bf{a}^l_{\rm{in}}$, $\bf{a}^r_{\rm{in}}$ (incoming waves
respectively from the left and right side -- see Figure~(\ref{model})),
 with the outgoing waves
$\bf{a}^l_{\rm{out}}$, $\bf{a}^r_{\rm{out}} $ in the following way:

\begin{equation} 
\{ \bf{a}^l_{\rm{out}}, \bf{a}^r_{\rm{out}} \} = 
S \{ \bf{a}^l_{\rm{in}}, \bf{a}^r_{\rm{in}} \}.
\label{asa}
\end{equation}

The scattering matrix $S$ has therefore the following block structure:

\begin{equation}
S = \left(
          \begin{array}{cc}
                \bf{r}  &  \bf{t}  \\
                \bf{t'} &  \bf{r'}  
            \end{array}
    \right).
\label{Smat}
\end{equation}

Matrices $\bf{r}$ i $\bf{r'}$ (reflection submatrices) and $\bf{t}$ and $\bf{t'}$ (transmission
submatrices) have size $M \times M$, where $M$ is the number
of open channels. Matrix $S$ has a block symmetry, when $\bf{r'}=\bf{r}$ and $\bf{t'}=\bf{t}$.
The expressions for the $S$-matrix elements have the following form \cite{EGST96}:

\begin{equation}
r_{nm} (E) = 
\frac{i}{\pi} 
\ \sum_{j,k=1}^{N}
[\Lambda(E)^{-1}]_{jk}
\ \frac{\sin(m y_j) \sin(n y_k)}{\sqrt{ k_m(E) k_n(E) }} \
\exp[ i(k_m x_j + k_n x_k)]
\label{rnm}
\end{equation}

and

\begin{equation}
t_{nm} (E) = \delta_{n,m} \ + \ 
\frac{i}{\pi} \
\sum_{j,k=1}^{N} [\Lambda(E)^{-1}]_{jk}
\ \frac{\sin(m y_j) \sin(n y_k)}{\sqrt{ k_m(E) k_n(E) }} \
\exp[ - i(k_m x_j - k_n x_k)] ,
\label{tnm}
\end{equation}

where $E=k^2$ is the energy of incident electrons. Elements of the  $N \times N$
matrix $\Lambda(E)$ read:

\begin{equation}
\Lambda_{jj} (E)  = \alpha + \frac{1}{\pi} \
\sum_{n=1}^{\infty} \left [
 \frac{1}{2n} - \frac{i\sin^2(n y_j)}{k_n(E)} \right ],
\label{lambdadiag}
\end{equation}

and

\begin{equation}
\Lambda_{jm} (E)  =  - \frac{i}{\pi} \ \sum_{n=1}^{\infty}
\frac{\exp( i k_n(E) |x_j - x_m|)}{k_n(E)} \sin(n y_j)
\sin(n y_m) , \qquad j \neq m.
\label{lambdaoff}
\end{equation}

The longitudinal momentum $k_n$ satisfies  the relation
\begin{equation}
k^2 = k_n^2 + n^2,
\label{klong}
\end{equation}
hence for $n > k$, it becomes imaginary $(k_n\sim i n)$,
what ensures the convergence of the series.

Conductance in the strip can be calculated using the Landauer formula:
\begin{equation}
G  = G_0 {\rm Tr} \{tt^{\dagger}\},
\label{landauer2}
\end{equation}
where $G_0= e^2/h$ (the spin degeneracy factor is omitted).

The total cross-section, $\sigma$, for the scattering on a single point--like impurity
can be evaluated and the result is following \cite{GSZZ98}
\begin{equation}
\sigma=\frac{\pi^2}{k}\frac{1}{[\gamma+\ln(k/2)]^2+\pi^2/4} ,
\label{cross}
\end{equation}
where $\gamma \approx 0.5772... \ $ is the Euler constant. 
The mean free path, $l_e$, can be calculated in a straightforward way (assuming the width
$W = \pi$):
\begin{equation}
l_e=1/\rho\sigma,  \qquad \rho=N/(L \pi).
\label{le}
\end{equation}

This parameter is important for determining various regimes of the scattering occuring
in a mesoscopic sample.
Thus choosing parameters of the model so as to keep the ratio of $N$ and $L$ constant
we are able to fix the mean free path for elastic scattering. 

\subsection{Statistical properties of the $S$-matrix ensemble}

For chaotic scattering models there is a conjecture that the Random Matrix Theory (RMT) 
correctly describes statistical properties of the $S$-matrix ensemble\cite{BS88,BM94}. 
Given that conjecture, the statistical properties of the unitary $S$-matrix can be described
by random matrices, which belong to appropriate Dyson circular ensembles \cite{B97}:
 Circular Orthogonal Ensemble (COE) in the presence of the Time Reversal Symmetry (TRS)
or Circular Unitary Ensemble (CUE), when the TRS is broken.
The RMT yields some predictions for the mean and variance
of the conductance, when there is TRS present and  no block symmetry (BS)
applied \cite{B97}:

\begin{equation}
\langle G \rangle_{\rm R} = \frac{M}{2} -
\frac{M}{4 M + 2},
\label{gavrmt}
\end{equation}

and

\begin{equation}
{\rm Var}_{\rm R} (G) =
\frac{M (M + 1)^2}{(2 M + 1)^2 (2 M +3)}.
\label{gvarmt}
\end{equation}
The above formulae yields
$\langle G \rangle_{\rm R} \approx 2.27 $
when the number of channels $M=5$ is taken.

Moreover, in the localization regime the 
following relation holds between variance of the
logarithmic conductance and the mean logarithmic conductance \cite{B97}: 
\begin{equation}
  {\rm Var} ( {\rm ln} G)  = 2 \langle - \ln G \rangle
\label{varlok}
\end{equation}

Another interesting prediction refers to the so called enhanced backscattering.
One may consider average 
over the COE or CUE ensemble of squared matrix $S$ elements, that is probabilities.
The prediction based on
the RMT yields ($\beta=1$ for COE and $\beta=2$ for CUE) \cite{B97}:

\begin{equation}
\langle |S_{mn}|^2 \rangle_{\beta} = \frac{1 - (1-2/\beta) 
\delta_{mn}}{2\ M -1 +2/\beta} ,
\label{scoue}
\end{equation}

In the presence of the TRS ($\beta=1$) the probability of scattering back to the same channel
is twice  the value of the probability of the scattering to a different channel.
When the TRS is broken ($\beta=2$) there is no longer enhancement in the backscattering 
-- the probability of backscattering is equally distributed over all channels.

\subsection{$S$ matrices with block symmetry}

For the ensemble of $S$-matrices with the block symmetry (BS) and TRS 
the RMT prediction yields \cite{KZ97}:

\begin{equation}
\langle G \rangle_{\rm R} = \frac{M}{2},
\label{gavrmtbs}
\end{equation}

and

\begin{equation}
{\rm Var}_{\rm R} (G) =
\frac{M }{8 + 4 (M - 1)}.
\label{gvarmtbs}
\end{equation}

The above formulae yields
$\langle G \rangle_{\rm R} = 2.5$ 
when the number of channels $M=5$ is taken and is slightly larger than the average
without the block symmetry (the difference is the so called weak localization correction,
which also vanishes when the TRS is broken).

\subsection{The Wigner--Smith time delay}

Some properties of the scattering are well described in terms of the Wigner--Smith
time delay \cite{W55,S60}. The time delay is related to the time spent in the scattering
region by a wavepacket whose energy is $E$. 
In the multichannel case, the Wigner-Smith time delays are
 defined in terms of the matrix $Q(E)$ \cite{LSSS95}, 
expressed in terms of the scattering matrix and its energy derivative:
\begin{equation}
Q(E) = -i S^{\dagger} \frac{dS}{dE}
\label{wsdelay}
\end{equation}
Eigenvalues of the matrix $Q(E)$, $\tau_i$, $i=1, \ldots 2M$, are called proper times of delay.

\section{RESULTS}

\begin{figure}[ht]
  \includegraphics[scale=0.45]{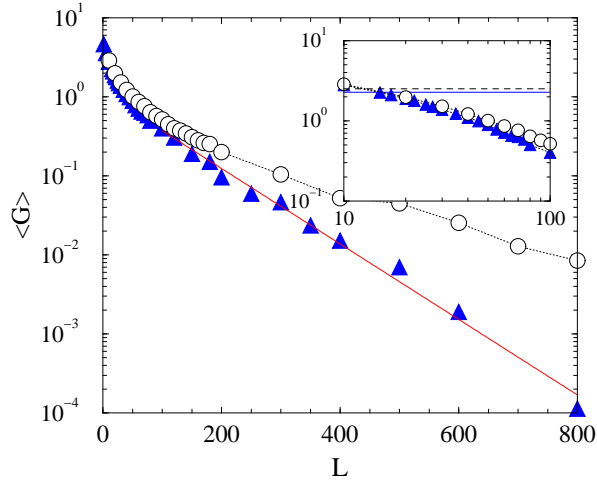} 
  \caption{
   Average conductance $\langle G \rangle$ in units of $G_0$ versus length of the wire $L$
   shown on a log scale. Tne number of scatterers is equal to the length, $N=L$
   This keeps a well defined mean free path.
   Tne number of channels $M=5$. Triangles show results for the model, with an ensemble of
   $500$ scattering matrices $S$, each corresponding to a different configuration of randomly
   distributed over the wire $N$ point scatterrers. Note, that for long samples the mean
   conductance decreases exponentially with $L$, $\langle G \rangle \sim \exp(- L/\xi)$.
   The localization length, $\xi$, has been found by the least--square fit, to have a value
   $\xi = 100 \pm 3$ for $L > 100$.
   Circles show results, when the
   scattering matrices have a block symmetry.
   The inset shows a double-log scale plot of the quasi-diffusive region, for $ 10 < L < 100$,
   for the $S$-matrices ensemble with and without block symmetry compared to the RMT predictions.
          }
  \label{avcond}
\end{figure}

Let us begin with demonstating various regimes of the electronic transport exhibited
by the model with the TRS present.
Figure (\ref{avcond}) shows on a log scale 
average conductance $\langle G \rangle$ in units of $G_0$ versus length of the wire $L$.
The width of the strip is $W=\pi$.
The wavenumber of the incident electron, $k$, is fixed at a value of $k=5.5708$. For that energy 
(which can be thought of as the Fermi energy in the contacts at temperature $T = 0$ Kelvin degrees)
there is $M=5$ open channels, and the corresponding $S$--matrix is 10 by 10.
The number of impurities $N=L$, so as to keep the constant mean free path $l_e \approx 9$.
The average has been computed out of 500 scattering matrices, each of them corresponding
to a random configuration of point--like scatterers.
For a long enough sample, i.e. $L > 100$,
(and correspondingly large number of impurities) one can observe that the
localization regime in transport sets in. This shows up in a characteristic exponential
behaviour of the mean conductance:

\begin{equation}
\langle G \rangle \sim \exp (- L/\xi) 
\label{gexp}
\end{equation}

The localization length, fitted to the data (represented by filled triangles)
yields a value $\xi = 100 \pm 3$.
The quality of the least--square fit shows the red line in the figure. 

\vskip 1.5truecm

In the same figure there are shown also results for the ensemble of $S$-matrices
with the block symmetry BS (open circles). 
The block symmetry is obtained, when the
distributed point scatterers have a mirror symmetry with respect to the $y$-axis (that is
$N/2$ points have been randomly distributed for $x \in (-L/2, 0)$ and $y \in (0, \pi)$ and
then the other $N/2$ have been obtained by the transformation
 $x \rightarrow -x, \quad y\rightarrow y$.
   
The inset in Figure~(\ref{avcond}) presents a double--log scale plot of the so called
``quasi--diffusive'' region, for $ 10 < L < 100$,
for the $S$-matrices ensemble with and without block symmetry compared to the RMT predictions.
 Note, that there is an interval, where the mean conductance is
roughly inversly proportional to $L$, $<G> \sim L^{-c}$. Fitted value $c = 0.90 \pm 0.05$ 
is close to 1.

\begin{figure}[ht]
  \includegraphics[scale=0.45]{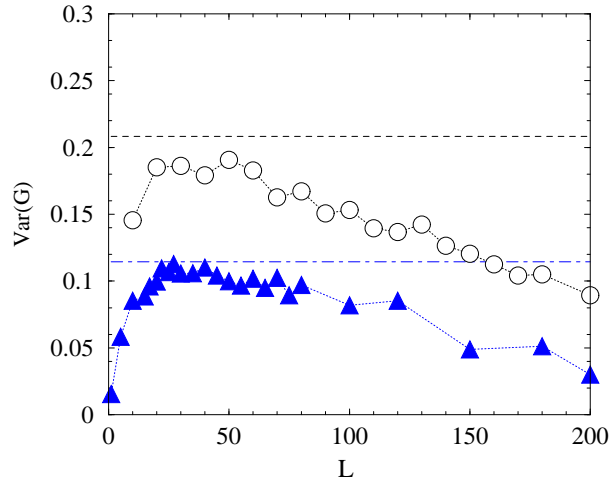} 
  \caption{
  Variance of the conductance Var(G) as a function of the sample size $L$.
  The TRS is present. Number of channels $M=5$. Triangles show results without
  $S$-matrix block symmetry, and circles with that symmetry imposed. Horizontal lines
  depict the RMT predicitions for both BS, and no BS present. 
          }
  \label{varcond}	  	  
\end{figure}

In order to illustrate further the intermediate, quasi-diffusive regime, let us
consider the variance of the conductance for discussed above ensembles of $S$--matrices.
Figure~(\ref{varcond}) shows dependence 
of the variance of the conductance on the sample size $L$.

Note that there is a narrow interval
in sample lengths $L \in (20, 50)$, where the variance weakly depends on the sample length. This
defines the intermediate (quasi-diffusive) transport regime, which separates regimes of 
the ballistic and localized transport.  
  
\begin{figure}[h]
  \includegraphics[scale=0.45]{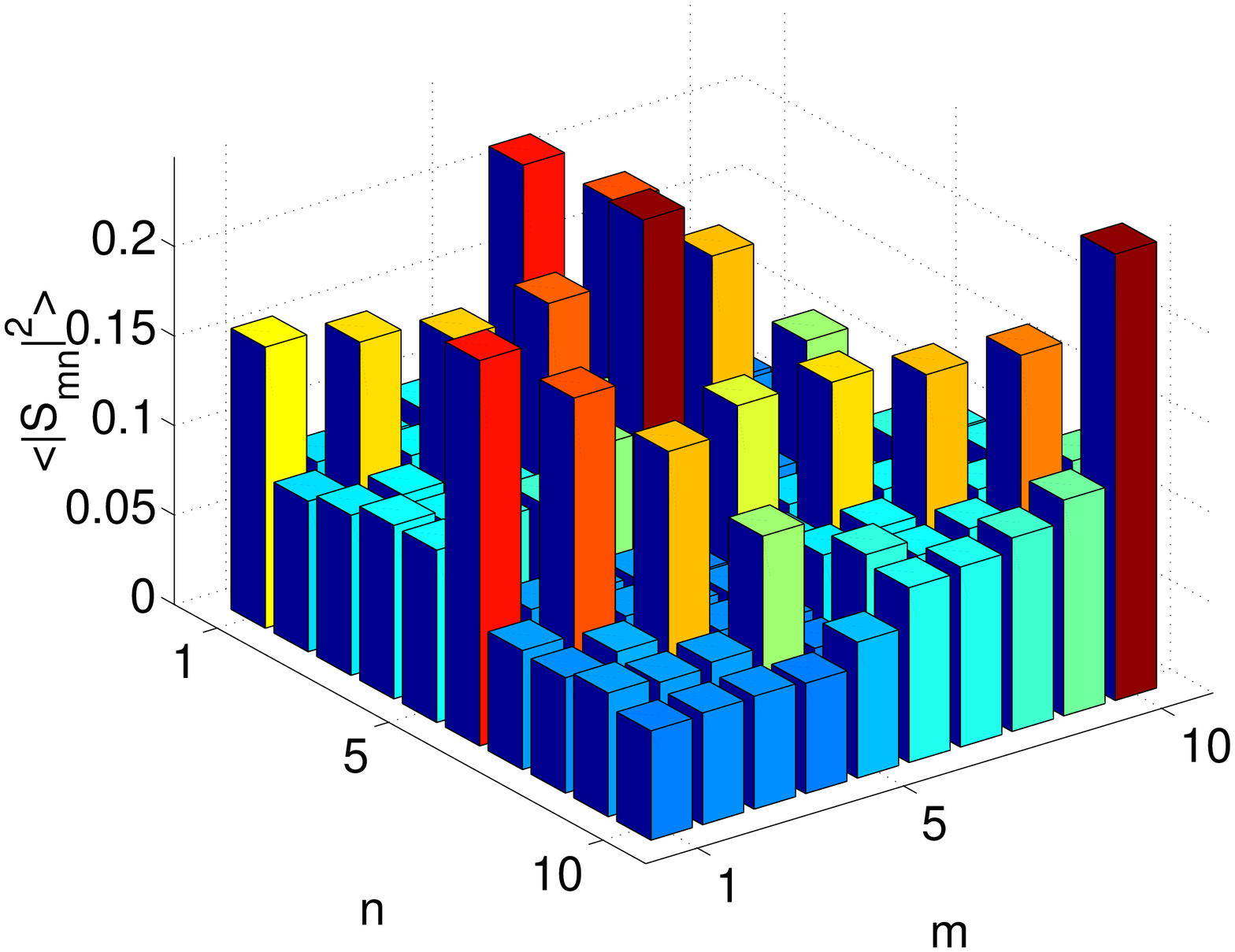} 
  \includegraphics[scale=0.45]{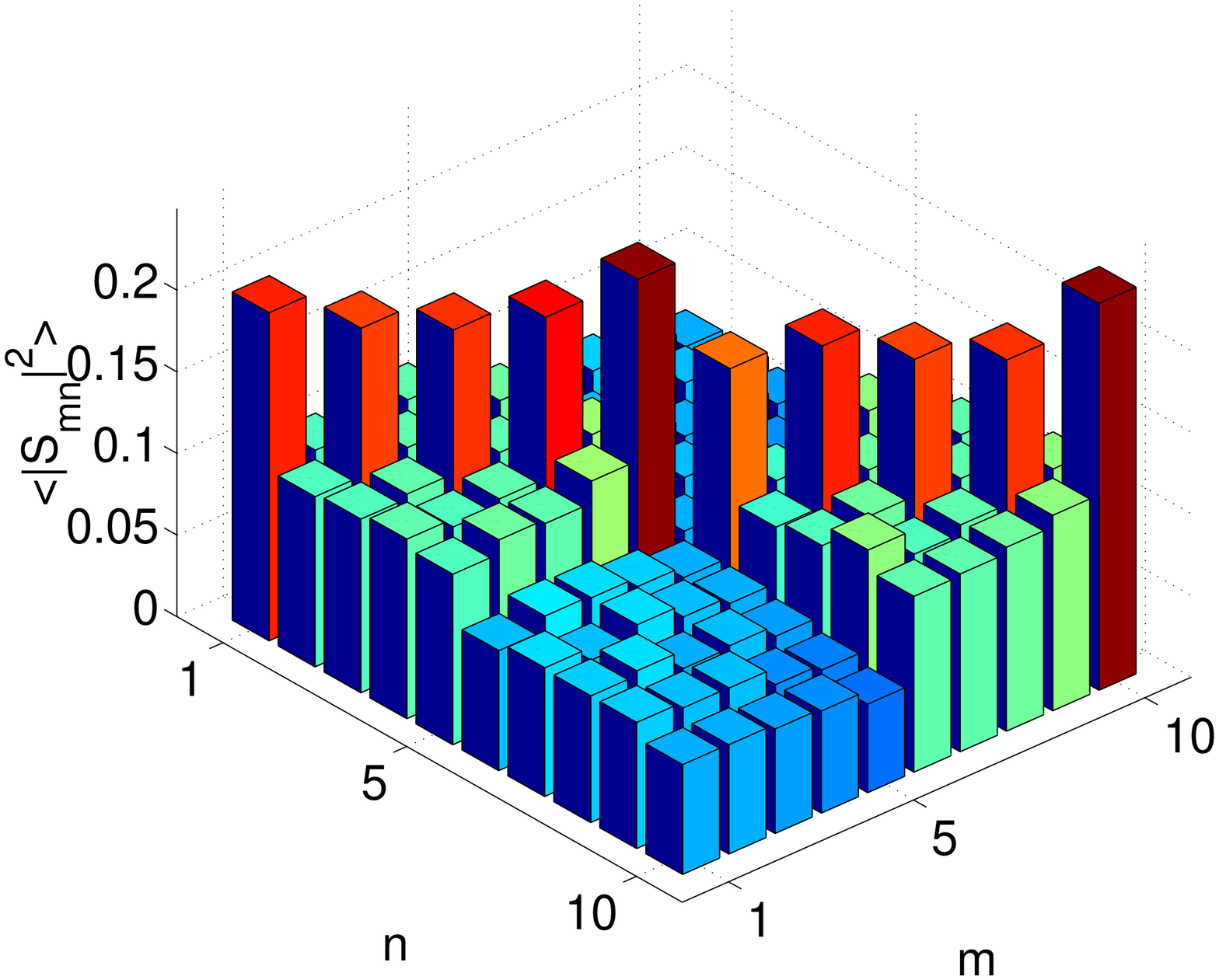}
\caption{
        Averege over ensemble of squared absolute values of the $S$-matrix elements.
        The left panel
	shows raw data, whereas the right panel presents the data after unfolding
	procedure. 
         Note, that after unfolding, it becomes clear that there exists enhancement
         on the diagonal.
	This indicates, that scattering back to the same channel is twice (roughly)
        larger than
	scattering back to a different channel. 
   }
 \label{3d}	  	  
\end{figure}

It is interesting to look into details of the $S$-matrix statistics in the
 intermediate regime, and
see if it bears any deeper resemblance to the diffusive regime than just the fact, that
the mean value of the conductance and its variance have close values to the ones predicted
 by the RMT.
Let us inspect Figure~(\ref{3d}).  

Figure~(\ref{3d}) presents 
averege  of squared absolute values of $S$-matrix elements evaluated for
an ensemble of 1000 $S$--matrices. In general however,
 an ensemble of $S$--matrices describing the scattering in the model wire 
 for a number of configuration of $N$ impurities
does not necessarily comply with the asumption for the RMT ensemble, saying that the average
of the matrix elements over the ensemble vanishes. This requirement is usually met in the pure
diffusive regime in the quantum chaotic scattering, where there is no direct processes present. 
 The left panel of the figure
shows raw data, whereas the right panel presents the data after unfolding procedure
described in \cite{FM85}. This procedure transforms the ``raw'' ensemble into
an equivalent ensemble, whose average over the ensemble vanishes
 $\langle S \rangle = 0$. It is clear, that after removing ``direct components'' (i.e. a memory of
 the incdent channel), there is much better agreement with Equation~(\ref{scoue}),
 describing an enhancement on the diagonal.
This indicates the fact, that scattering back to the incident channel
is indeed roughly twice as large than
scattering back to a different channel. Therefore in the present model we recover a well known
universal result.

\begin{figure}[h]
  \includegraphics[scale=0.45]{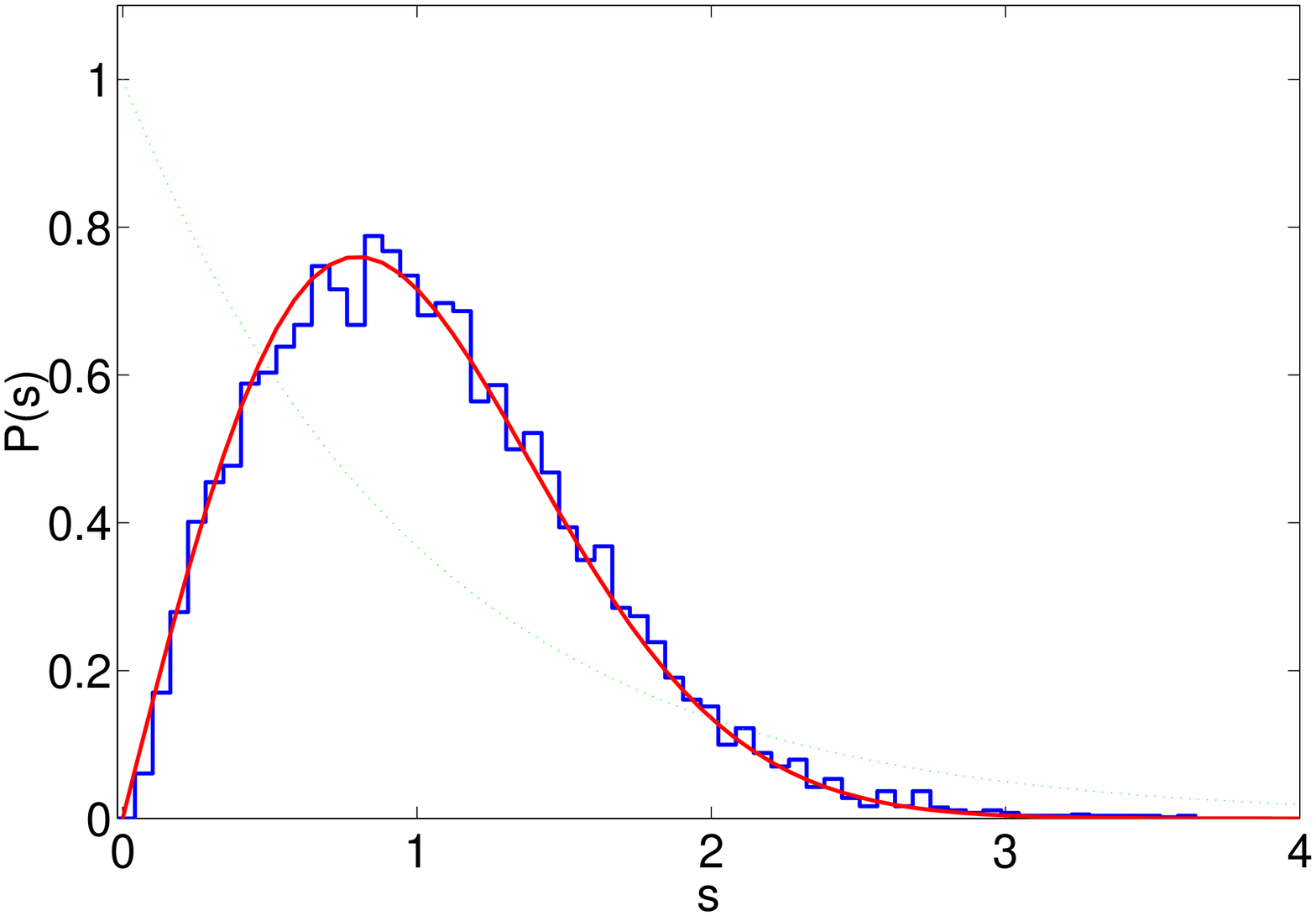}
\includegraphics[scale=0.45]{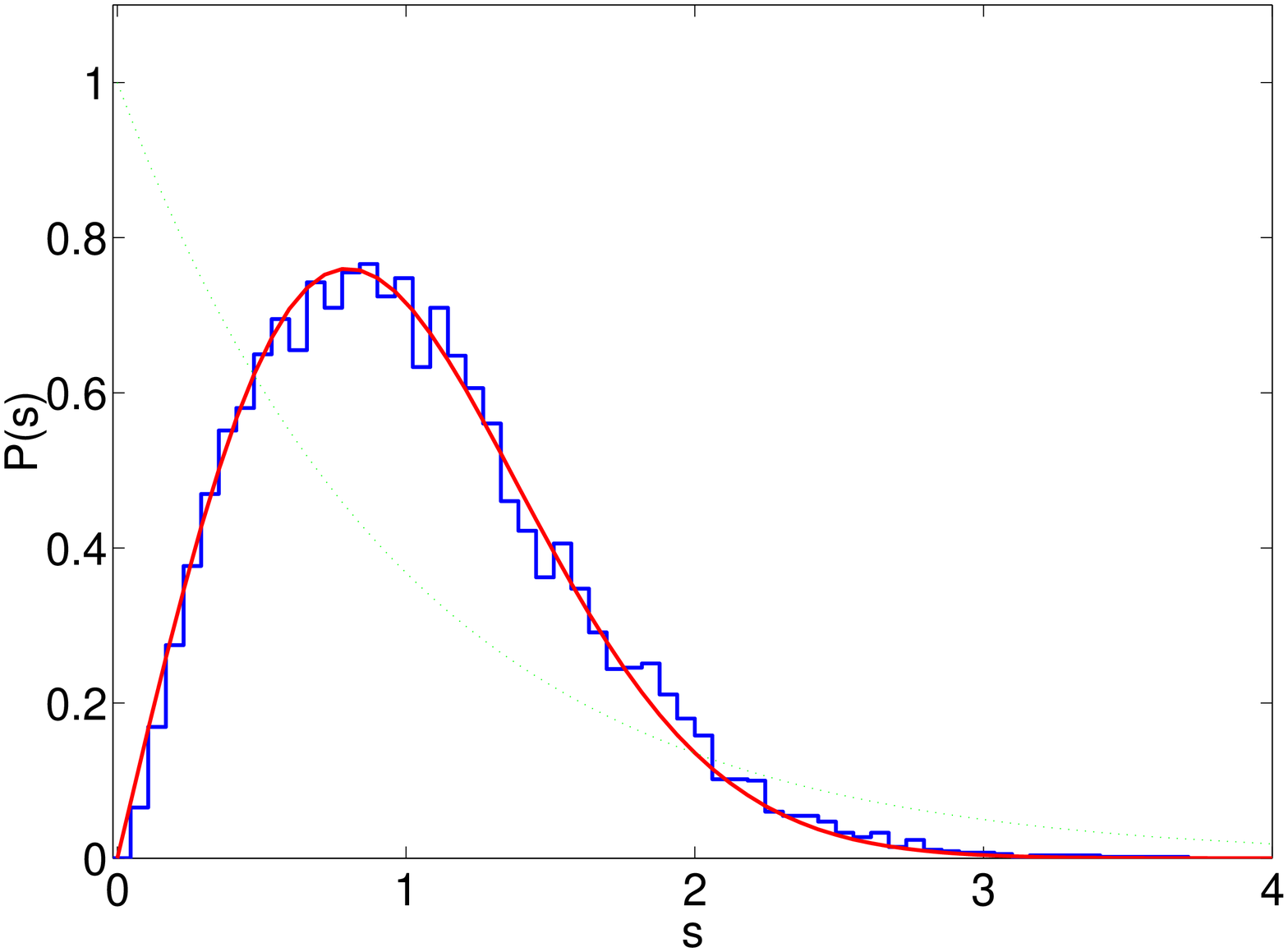}
	 \caption{
	 Nearest neigbour statistics, $P(s)$, for raw data (left panel) and unfolded
	 data (right panel). Parameters the same as in 
         { \protect{Figure~(\ref{3d})} }.
	 Note that there is virtually no difference between results shown on both panels.
         Both panels show a good agreement with the Wigner surmise (red curve).
	 For a reference, there is also shown a Poisson distribution 
	 $P(s) = \exp(-s)$ (green curve). 
	         } 
  \label{psunfo} 
\end{figure}

It is interesting to note, that the nearest neighbour NNS 
statistics, both for raw $S$-matrix eigenphases and unfolded data
exhibit the Wigner behaviour --- see Figure~(\ref{psunfo}).

\begin{equation}
 P(s) = \frac{s\pi}{2} \exp \left[-\frac{s^2\pi}{4} \right],
\label{GOE}
\end{equation}

where $s$ denotes the spacing of the $S$-matrix eigenphases.
This means that the NNS statistics, $P(s)$, is not so sensitive in revealing the universality
regime, which complies fully with the RMT requirements and predicitons.
 
\vskip 0.5truecm 

In summary, the ``quasi--diffusive'' regime disscussed above bears some
resemblance to the universal
diffusive regime at the microscopic level, where the standard RMT theory of transport 
\cite{B97} holds, provided
that one looks at statistical properties of the fluctuations around mean values of the
$S$-matrix ensemble rather that considers the ``raw'' ensemble itself.  
At the  macroscopic level, the conductance shows up some indication
of the ``ohmic behaviour'', that is $\langle G \rangle \sim 1/L$, and the variance
of the conductance is weakly dependent on the sample size. The ``quasi--diffusive''
regime of transport is a transition regime between the ballistic transport and the 

\begin{figure}[h]
\includegraphics[scale=0.45]{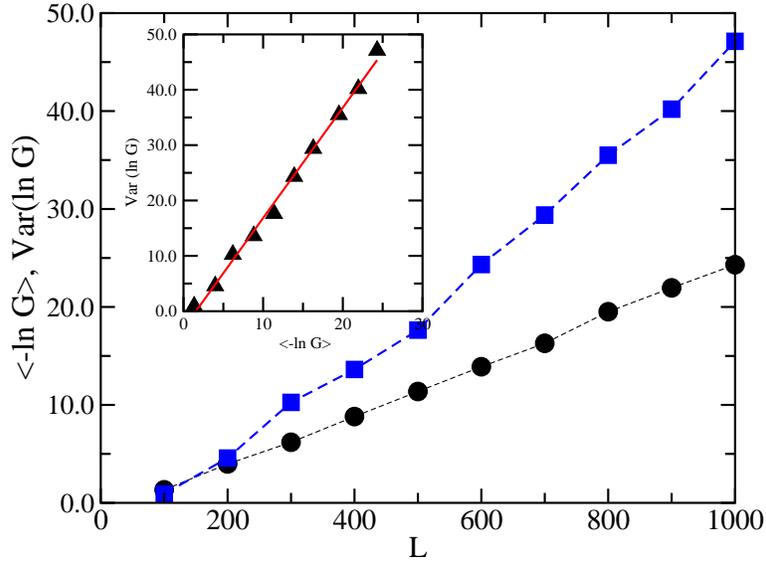}
\caption{
      Average logarithmic conductance (filled dots) and variance of the logarithmic conductance
      (filled squares)
      as a function of the sample length. The inset ilustrates corelation
      between average and the variance of the logarithmic conductance ${\rm ln} G$.
      Each point corresponds to a different sample length $L$.
      The inset shows correlation between
      the variance and the average of the ${\rm ln} G$.
      } 
\label{lokal}
\end{figure}

In Figure~(\ref{avcond}) we have seen that for $L > 100$ one can see exponential decay
of averaged over ensemble conductance. Instead of conductance itself let us look at
logarithm of conductance. Figure~(\ref{lokal}) presents both average (filled dots)
and variance (filled squares)
of the logarithmic conductance.
The inset shows the corelation between variance and the average
of $ {\rm ln} G$. The fitted slope $1.99 \pm 0.05$ is in perfect agreement with the theoretical 
prediction of 2. This result supports a claim, that this regime is indeed 
characterized by the localization
in electronic transport through the model quantum wire. Again considered model captures well
features of the Anderson localization.

\begin{figure}[h]
\includegraphics[scale=0.45]{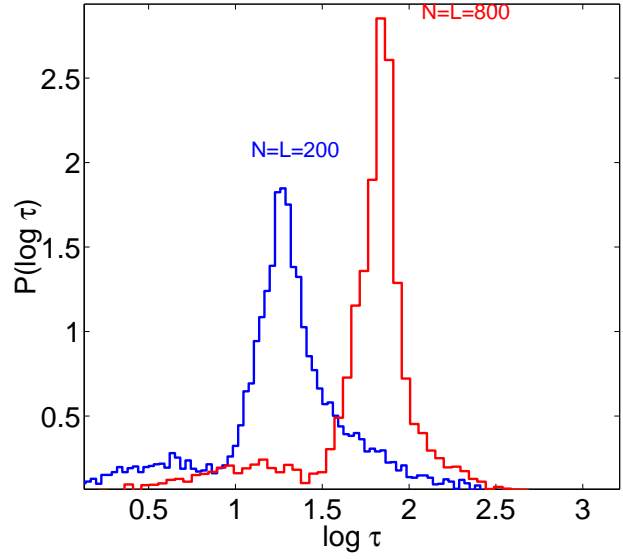}
 \caption{
	  Distribution of logarithms of time delays, ${\rm log} \tau$ for $L=N=200$ and $N=L=800$.
	         } 
  \label{tdelog}	  	  
\end{figure}

In order to see some more details of the transport mechanism, let us consider time
delay distributions in the region of the localized transport.
Figure~(\ref{tdelog}) shows a distribution of logarithms of
of the time delay, $\log \tau$ for $L=200$ and $L=800$. For longer samples ($L=800$),
the distribution shows well pronounced maximum for time delays that are one order
of magnitude longer than in the case of shorter samples ($L=200$).

\begin{figure}[h]
\includegraphics[scale=0.45]{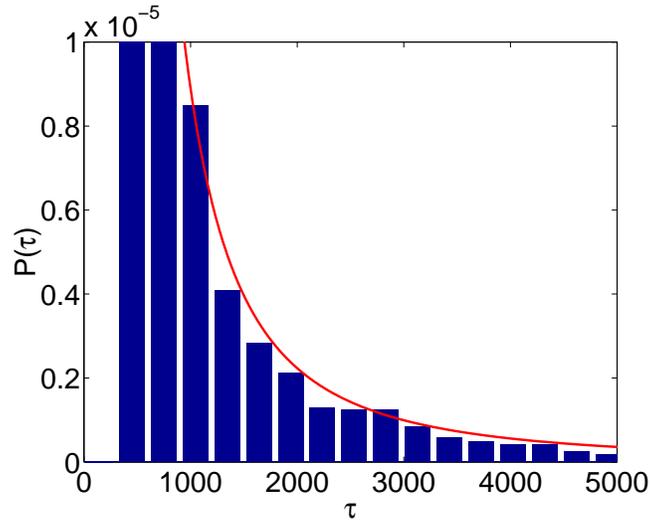}
 \caption{
	  Time delay distribution $P(\tau)$ for $L=N=400$. Note a good
	  agreement with the prediction 
          for a one-dimensional random potential in the tail of
	  the distribution (solid curve).
	  There is $10^4$ $S$-matrices in the ensemble to calculate this distribution.
	         } 
  \label{tdelay} 
\end{figure}

Figure~(\ref{tdelay}) presents the tail of the distribution
$P(\tau)$ of time delays obtained for an ensemble of $10^4$ scattering matrices. 
The distribution resulting from adopted here theoretical model is compared with 
a theoretical prediction of \cite{TC99}.
The fitted curve on the right panel of Figure~(\ref{tdelay}) has been
derived for a one-dimensional random potential in \cite{TC99} and reads:
\begin{equation}
P(\tau) = \frac{\xi}{2 k \tau^2} \exp \left [ - \frac{\xi}{2 k \tau} \right ]
\label{ptau}
\end{equation}
The localization length $\xi$ is taken to have a value of 100 (as computed above) and $k=5.5708$. 
Thus presented here two-dimensional model essentially behaves like one-dimensional model
in the limit of long times. Therefore also with this respect, the model exhibits a generic behaviour.

\section{SUMMARY}

Presented continuous and solvable
(though not in a form of closed analytical expressions)
model of electronic transport in disordered quantum wires
is capable of reproducing a whole range of universal
phenomena. 
Both micro ($S$-matrix statistics) and macro (conductance)
signatures show consistant features of
the ballistic, intermediate ``quasi-diffusive'' and Anderson localization regime
in the transport. 
Results of the model agree very well with the predictions of the standard
transport theory based on the RMT for a wide range of parameters.
This allows to identify regimes of universal behaviour of the system.
Moreover, thanks to flexibility and simplicity, presented model offers interesting
possibilities of extending the standard transport approach beyond
the quasi one--dimensional case in the regime of diffusive scattering.

\section*{ACKNOWLEDGMENTS}
The author wishes to thank the Organizers for financial support of his 
attendance to the Nanotubes and 
Nanostructures 2001 School and Workshop in Frascati, Italy.

%
\end{document}